%% file: template_arXiv.tex
\documentclass{article}

\usepackage{arxiv}

\usepackage[utf8]{inputenc} 
\usepackage[T1]{fontenc}    
\usepackage{hyperref}       
\usepackage{url}            
\usepackage{booktabs}       
\usepackage{amsfonts}       
\usepackage{nicefrac}       
\usepackage{microtype}      
\usepackage{lipsum}		
\usepackage{natbib}
\usepackage{doi}
\input{lib/preamble.tex}

\usepackage{algorithm}
\usepackage{algpseudocode}
\usepackage{listings} 

\lstdefinestyle{mystyle}{
  basicstyle=\ttfamily\footnotesize,
  frame=single  
}
\title{PESC – Parallel Experiment for Sequential Code}

\author{ 
        \href{https://orcid.org/0000-0002-9544-7774}{\includegraphics[scale=0.06]{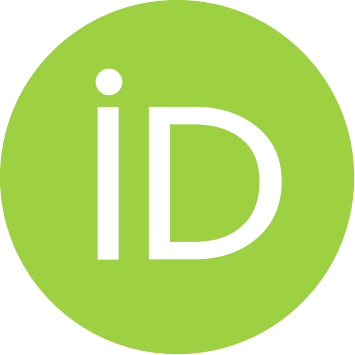}\hspace{1mm}Henrique C. T. Santos}\thanks{95, Av. Prof. Luís Freire, 500 - Cidade  Universitária, Recife - PE} \\
	Instituto Federal de Pernambuco\\
	Recife, Brasil \\
	\texttt{henrique.santos@recife.ifpe.edu.br} \\
        \And
        \href{https://orcid.org/0000-0002-6527-7065}{\includegraphics[scale=0.06]{orcid.pdf}\hspace{1mm}Luciano S. de Souza}\thanks{95, R. Manuel de Medeiros, 35 - Dois Irmãos, Recife - PE} \\
	Departamento de Estat\'{i}stica e Inform\'{a}tica\\
	Universidade Federal Rural de Pernambuco\\
	Recife, Brasil \\
	\texttt{luciano.serafim@ufrpe.br} \\
	\And
	\href{https://orcid.org/0000-0002-2672-7801}{\includegraphics[scale=0.06]{orcid.pdf}\hspace{1mm}Jonathan H. A. de Carvalho} \\
	Centro de Inform\'{a}tica\\
	Universidade Federal de Pernambuco\\
	Recife, Brasil \\
	\texttt{jhac@cin.ufpe.br} \\
	\And
	\href{https://orcid.org/0000-0002-2131-9825}{\includegraphics[scale=0.06]{orcid.pdf}\hspace{1mm}Tiago A. E. Ferreira} \\
	Departamento de Estat\'{i}stica e Inform\'{a}tica\\
	Universidade Federal Rural de Pernambuco\\
	Recife, Brasil \\
	\texttt{tiago.espinola@ufrpe.br} \\
}

\renewcommand{\shorttitle}{PESC - Parallel Experiment for Sequential Code}

\hypersetup{
pdftitle={A template for the arxiv style},
pdfsubject={q-bio.NC, q-bio.QM},
pdfauthor={David S.~Hippocampus, Elias D.~Striatum},
pdfkeywords={First keyword, Second keyword, More},
}

\begin{document}
\maketitle

\input{sections/abstract}
\input{sections/introduction}
\input{sections/related_works}
\input{sections/system_overview}
\input{sections/architecture}
\input{sections/evaluation}
\input{sections/experimental_results}
\input{sections/conclusions}

\input{template_arXiv.bbl}
\bibliographystyle{unsrtnat}






\end{document}

%% file: lib/preamble.tex
\usepackage{braket}
\usepackage{graphicx}
\usepackage{amsmath}
\usepackage{amssymb}
\usepackage{amsfonts}
\usepackage{braket}
\usepackage{float}
\usepackage[table,xcdraw]{xcolor}
\usepackage{cite}
\usepackage{subfig}

\usepackage{setspace}
\usepackage{multirow}

%% file: sections/abstract.tex
\begin{abstract}

The need for computational resources grows as computational algorithms gain popularity in different sectors of the scientific community. This search has stimulated the development of several cloud platforms that abstract the complexity of computational infrastructure. Unfortunately, the cost of accessing these resources could leave out various studies that could be carried by a simpler infrastructure. In this article, we present a platform for distributing computer simulations on resources available on a network using containers that abstracts the complexity needed to configure these execution environments and allows any user can benefit from this infrastructure. Simulations could be developed in any programming language (like \verb|Python|, \verb|Java|, \verb|C|, \verb|R|) and with specific execution needs within reach of the scientific community in a general way. We will present results obtained in running simulations that required more than 1000 runs with different initial parameters and various other experiments that benefited from using the platform.

\end{abstract}

%% file: sections/introduction.tex
\section{Introduction}
\label{sec:introduction}

With the creation of public cloud services such as Amazon AWS \citep{aws:2022}, Microsoft Azure \citep{azure:2022}, and Google Cloud \citep{google_cloud:2022}, Desktop Grid solutions \citep{choi2007characterizing} were gradually being replaced by solutions on these platforms or extended using these features \citep{kravsovec2019enhancing}. A reason for this migration is the high cost of acquiring and maintaining the local computational infrastructure, and the information technology (IT) team training \citep{9383971}. However, despite offering free plans, these public cloud services have access restrictions, not favoring the time-consuming simulations or simulations that require specialized resources \citep{GoogleFAQ:2022}. Thus, accessing the non-free resources offered by these platforms may be impossible for many institutions, like public educational institutions in developing countries \citep{NJENGA2019225}. 

However, even in developing countries, educational and research institutions always have computer laboratories, where desktop computers are used for study activities, teaching in the classroom, and research experiments. Therefore, the institution could use all this equipment to generate a local infrastructure to create a computational environment capable of running time-consuming simulations. Nevertheless, the big problem in this situation is configuring and supporting the computational environment, so it may be necessary to train the IT team. At the same time, even with a computational infrastructure set up and configured, the users will also need to learn computational parallel and high-performance techniques to access the new environment's computational power. 

Inspired by the presented scenario, we propose a way to take advantage of existing infrastructure by developing a computational platform that combines all available computing power, allowing the current IT professionals to keep a solution as a local service, easy to maintain, and configure. Furthermore, it enables the academic community to run codes and simulations of their research or use it as a didactic tool in teaching parallel programming. The Parallel Experiment for Sequential Code (PESC) is a platform that aims to allow access to idle computing resources available. This platform will enable users to configure their execution environments and run codes and computer simulations on these resources. All user interaction with the PESC is carried out through a web interface, guarantying autonomy to users both in the execution process and in the monitoring of requests in progress. Each user code associated with an execution request will be packaged in a Docker \citep{merkel2014docker} container and distributed by a server to the client computers that will execute the code. In addition, the server will monitor the status of each execution to redistribute the workload and perform reallocations to other computers in case of failure or lack of communication. Due to the use of containers, programs created in different programming languages can be executed on the PESC platform.

For this purpose, platforms, frameworks, and tools have emerged in recent decades, such as CONDOR \citep{litzkow1987condor} and BOINC \citep{anderson2004boinc}. They leverage idle resources on computers for the distribution of distributed programming tasks. The difference between them is that CONDOR works with resources from a private grid and BOINC with voluntary computing through access to public computers. Other works have been developed more recently, such as Everest \citep{closer14}, OSCAR \citep{perez2019premises}, DLHub \citep{li2021dlhub}, and ARC \citep{prila2021development}. Although they can be employed in the context of the problem presented in this article, they are focused on solving specific problems, interfere with user code, and/or do not give a user autonomy to use the platform. Furthermore, it could generate a demand for infrastructure administration that a reduced IT staff might not be able to handle.

Another characteristic of PESC is that it is minimally intrusive in user code, allowing legacy programs to benefit from the platform without making significant changes to their code. The no intrusion is because it is not simple to parallelize programs initially written to run sequentially \citep{Bhalla2014}. In this sense, PESC simplifies this process through parameters that identify the instance currently running, allowing sequential codes that use independent loops to be easily adjusted to run in parallel, where each iteration of the loop will be running in a different container.

We evaluate the PESC platform in controlled and realistic situations using machines with configurations similar to those found in laboratories of educational institutions. We show that the PESC platform speeded up the execution of sequence codes with substantial time savings, even without (or minimal) modification to the user code.

The rest of this work is structured as follows. In Section \ref{sec:related_work}, we describe existing computational solutions for the parallel execution distribution. Next, the Section \ref{sec:system_overview} presents an overview of the platform. The Section \ref{sec:architecture} presents the PESC architecture and how its components communicate. In Section~\ref{sec:evaluation}, we evaluate the platform through controlled tests. In Section \ref{sec:experimental_results}, we offer real use cases that employed the platform and benefited from the proposal presented. Finally, the Section \ref{sec:conclusions} concludes the work shown.

%% file: sections/related_works.tex
\section{Related Works}
\label{sec:related_work}

Several studies on desktop grids were carried out, some of which were analyzed in review works \citep{Khan2017,ivashko2018survey}. Nowadays, not all the solutions reviewed in these survey publications remain active. Besides, almost all analyzed solutions do not adhere to the proposal presented here, where some require specific libraries need to be installed for the execution of the user code \citep{htcondor}, or they are built for a particular programming language \citep{almeida2019general}, or the user must rewrite the code based on the requirements defined by the project \citep{aldinucci2021practical}. With the evolution of technology, it is possible to rethink how these desktop grids can be implemented, focusing on user-friendly adoption, portability, and no (or minimum) user-code intrusion to run the computational experiments.

Several available technology options allow configuring an environment to distribute the running process load on different network computers, such as Docker Swarm \citep{DockerSwarm:2022} and Kubernetes \citep{kubernetes:2022}. Many of these options depend on a specialized IT team setting up these environments or are created to provide services in cloud infrastructure. For example, web servers or database servers as presented in~\citep{DockerSwarm:2022} and \citep{kubernetes:2022}.

One of the bases for the solution to this type of problem is presented in \citep{litzkow1987condor}, which has evolved into a more robust and updated tool called HTCONDR. Despite its simplicity, the user who adopts this solution must know how to use the command line tools and learn a task description language to submit their codes. The PESC platform is built in a way that gives autonomy to a user without specific training in the area of computer science and can request the execution of their codes through an intuitive web interface.

Another example is the BOINC \citep{boinc:2022}. Developed at the University of California, Berkeley, the BOINC is a platform that allows voluntary computing and grid computing in the local environment. In a local grid computing environment, the  BOINC also uses idle features of existing computers to perform works distributed by a server. The user must create a project for every need and configure their code as a BOINC app. The PESC platform shares some grid computing concepts in the local environment of BOINC. Still,  the PESC abstracts the complexity of the platform to make its adoption easier so that any user can run their codes and simulations.

Another platform presented for this purpose is EVEREST \citep{closer14}, where a set of computers can be registered to a server, centralized and managed by the creators of the platform, and start to receive tasks for execution in a distributed way. The PESC platform is entirely contained in the institution's dependencies, guaranteeing secrecy and protection for sensitive studies. Furthermore, because it is based on containers to create the execution environments of the programs, it does not require the installation of dependencies related to the user code on the computers that will run it.

Other solutions are based on technologies related to cloud infrastructures, such as Serverless and Functions as Services \citep{hassan2021survey}. Among them, we can highlight OSCAR \citep{perez2019premises}, which presents a framework to efficiently support Functions as a Service in a local environment for general-purpose file-processing computing applications. The PESC platform minimizes using other products or frameworks to create an ecosystem simpler for IT staff to manage and is designed to run general-purpose scientific applications.

Other more recent works have solutions for specific problems, but given the nature of their architecture, they can also be used for the distribution of works on others computers. For example,  DLHub \citep{li2021dlhub} is a learning system that provides template publishing and service capabilities for scientific machine learning. The service infrastructure is based on funcX \citep{10.1145/3369583.3392683}, a function as a service platform developed specifically to support remote and distributed functions execution. The PESC platform has similar goals but is designed in a way that does not determine how the user code should be written, allowing legacy programs to run without modifications.

%% file: sections/system_overview.tex
\section{System Overview}
\label{sec:system_overview}

In this section, we introduce the PESC platform describing its main components and making an overview of the system workflow.

The objective of the PESC platform is to allow its users to run computer programs with idle computational resources. The user does not need to perform specific settings in the computational framework or does not make significant changes in their programs, and in many situations, no code modification. Furthermore, the user programs can run several times (for example, the statistical simulation needs many repetitions), in parallel or not, in a transparent way for the user. The PESC is accessed via an intuitive web interface so users can upload and run their code.

\begin{figure*}[!tb]
    \centering
    \includegraphics[width=\linewidth]{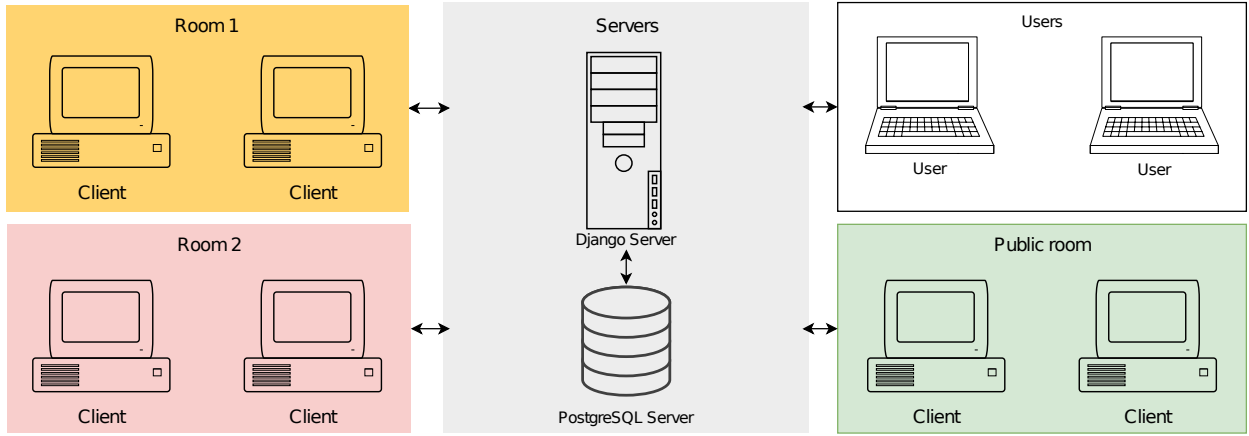}
    \caption{PESC system overview}
    \label{fig:overview}
\end{figure*}

Here, the idle computing resources will be called \verb|clients|. \verb|Rooms| are used to group \verb|clients| available on the PESC platform. These \verb|rooms| can represent the physical allocation of the \verb|clients| in the user environment, generating work modules, as described in Figure~\ref{fig:overview}. When a \verb|client| connects to the platform for the first time, it will only be visible to the administrator user. The administrator user can add this client to a public room, where all users will have access to it, or in a more restricted access room, where a user manages who will have access to it. Any user can create rooms, and once the administrator allocates clients to one of these rooms, this user will be able to reallocate clients between them.

The PESC platform is a non-intrusive solution to maintain compatibility with existing computer programs. It allows its users to run their codes without significant modifications (or, in many cases, without changes). The user code will run as a command line call with the parameters related to the current execution. Therefore, the user code must include a header to use those parameter values sent at the execution time but can ignore this header and run the code without knowing those values. The PESC platform provides the header code on its site, simplifying this configuration procedure for the user. It is worth emphasizing that the header is already available for different programming languages, such as \verb|Python|, \verb|Java|, \verb|C/C++|, and \verb|R project|, among others. If the header is not in the user code, then the PESC could include it automatically.

The header defines default values and will not interfere with executing the code outside the PESC platform. The list of parameters defined by the header is listed below:

\begin{itemize}    
    \setlength\itemsep{-0.2em}
    \item \verb|app|\_\verb|dir|: The current directory where the \verb|client| executes the code;
    \item \verb|checkpoint|\_\verb|dir|: The directory where crash recovery points are to be stored;
    \item \verb|output|\_\verb|dir|: The directory that will return to the server after executing the code. The user must store all the desired output of their code in this directory;
    \item \verb|rank|: Each instance of the user code receives an identity value which is represented by a value starting at zero; 
    \item \verb|repetitions|: The number of code run repetitions requested by the user;
    \item \verb|master|\_\verb|addr|: The IP address of the \verb|client| that is running the instance with \verb|rank = 0| of the current user code; 
    \item \verb|master|\_\verb|port|: The port associated with the IP address of the \verb|client| that is running the instance with \verb|rank = 0| of the current user code;
    \item \verb|parameters|: A user-defined list of values on the execution request form received by the user code as a vector. 
\end{itemize}

The PESC platform executes the user code in a Docker container. To configure the container execution environment, the user must create a \verb|Dockerfile| and a \verb|requirements.txt| file. The \verb|Dockerfile| defines the container environment defining operational system dependencies, and the \verb|requirements.txt| describes a list of libraries used to customize the \verb|Python| environment, the PESC platform base language. The user itself can provide these files to build the container image, but to keep the platform simple, the user can use a \verb|store| with a set of commonly used technologies. If no option in the \verb|store| meets the user's needs, the user can request it to the platform. This request can be answered by the system administrator or by any other user of the platform. However, if simple users attend the request, it will be available to requesting user after the validation by users authorized by the system (like a moderator). In the PESC platform, all that execution environment configuration is called \verb|domain|.

After defining the \verb|domain|, the user must configure the \verb|process| that the PESC platform will execute. The \verb|process| is the user code files that will run in the containers and can be provided as a single file in a given programming language, like \verb|Python|, or a Zip file if it is a code in multiple files. Finally, the server distributes the container definition to the \verb|clients|, which will control the entire build flow of dockers' images and the container's lifecycle execution associated with the user code.

The PESC platform also allows users to upload files as \verb|shared files| that the \verb|clients| can request at the execution time of the user \verb|process|. These files remain shared resources for all instances of the same user's \verb|processes| running on that \verb|client|. This share eliminates the need to transfer the same file to each instance of the same \verb|process|. Further, all these \verb|shared files| will have a read-only restriction. Thus, those share files also do not affect the container functioning, avoiding race conditions. An example of this resource use is a database in \verb|CSV| file format, which will be loaded during the execution of the user \verb|process|. For example, in a given execution, the user requests to execute their code a given quantity of times in a \verb|room| with two or more \verb|clients|. Each \verb|client| will receive only one copy of the database file, which will be available for each \verb|process| instance execution.

\begin{figure}[ht]
    \centering
    \frame{\includegraphics[scale=1.2]{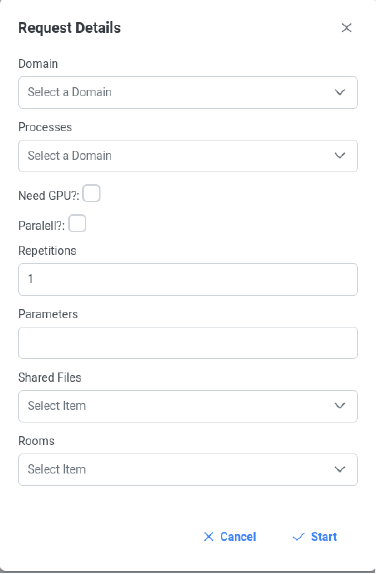}}
    \caption{The request form.}
    \label{fig:test2_request}
\end{figure}

After the \verb|domain| and \verb|process| configuration and uploading \verb|shared files|, if necessary, the user must make an execution request, which the PESC platform will call only the \verb|request|. The platform adds the \verb|request| to the user execution queue and distributes it when \verb|clients| have available resources. When creating the \verb|request|, the user must inform the data,

\begin{itemize}
    \setlength\itemsep{-0.2em}
    \item Domain: The definition of the execution environment;
    \item Process: The user code to be executed;
    \item Repetitions: The number of times the platform will execute the \verb|process|. The \verb|clients| will perform each execution in a different container; 
    \item Parallel: Whether the code makes use of parallel resources. This value will inform \verb|clients| to wait for the distribution of all requested copies before starting execution; 
    \item Parameters: A comma-separated list of values that will be received as parameters by the \verb|process| instances at run time; 
    \item GPU: Whether the \verb|process| needs GPU. This value indicates that only \verb|clients| that have this characteristic should be selected; 
    \item Same machine: The user must inform if all \verb|process| instances must run on the same \verb|client|. The user code should make use of sharing local resources such as memory or GPU; 
    \item Shared files: Files that are needed to run the \verb|process|. The user must have previously uploaded these files; 
    \item Rooms: Groups of \verb|clients| should receive the \verb|process| to execute.
\end{itemize}
Figure \ref{fig:test2_request} shows how this data is informed on the web interface PESC platform.

\begin{figure}[ht]
    \centering
    \frame{\includegraphics[scale=2]{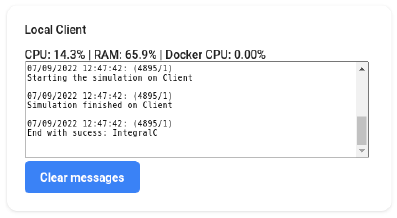}}
    \caption{Client Status Panel}
    \label{fig:client_status_panel}
\end{figure}

The \verb|request| execution monitoring can be performed by the user, in real-time, through a \verb|client| status panel available in the web interface, as shown in Figure \ref{fig:client_status_panel}. The monitor only displays the start and end of execution messages. However, if it is a process integrated with the resources of the PESC platform, the monitor can send execution details, such as custom messages and execution percentages. The user must implement a \verb|Python| class with specific definitions to integrate the code with the platform resources. A model of this class is available in the platform's web interface. However, it is worth noting that this integration with the platform is not an obligatory requirement to run the user's computational experiments.

Following the PESC sequence flow, the \verb|clients| receive the user \verb|process| instances defined in the \verb|request| and verify if has already built the Docker image for this \verb|process|. In the negative case, it starts the image-building process. Before the container execution, the \verb|client| checks if the \verb|shared files| configured in the \verb|request| are already received and request them to the server in the negative case. After that, the proper computational experiment starts. 

During the \verb|request| execution time, the client periodically checks with the server if the user canceled it. If the user cancels a \verb|request|, the \verb|client| interrupts the execution of all \verb|process| instances associated with it. After the execution (the user did not cancel the request), the \verb|client| compacts the output directory and sends it back to the server. In this directory, an \verb|output.txt| file is created with all the screen outputs performed by the program. For example, in the case of \verb|Python| programs, the results of the \verb|print| or \verb|printf| commands.

\begin{figure*}[!tb]
    \centering
    \includegraphics[width=\linewidth]{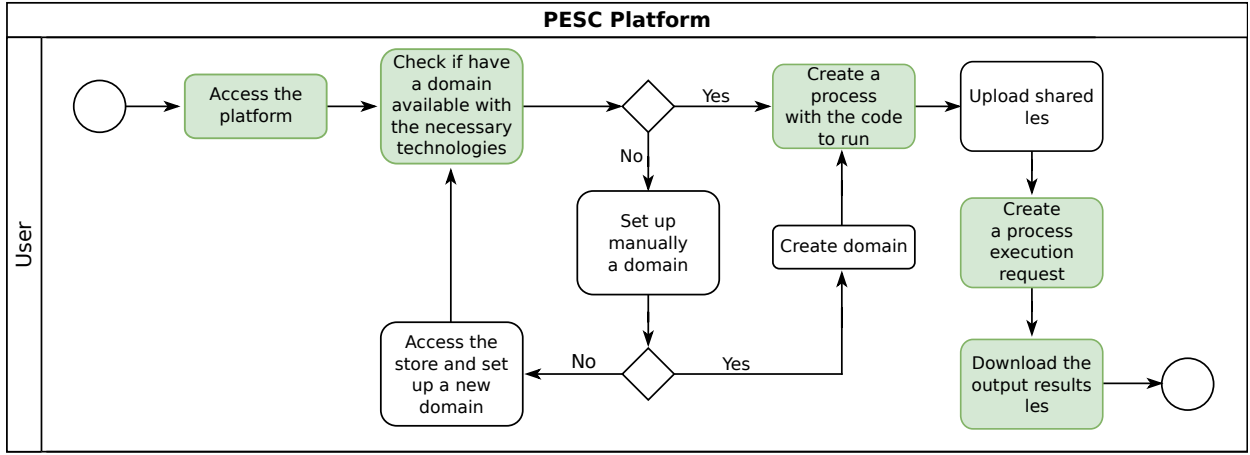}
    \caption{PESC Platform Flow. The elements in green represent the steps with mandatory user interaction.}
    \label{fig:overview_flow}
\end{figure*}

After the execution of the \verb|request|, the user can start downloading all the output files created by each \verb|client| involved in running this \verb|request|. Finally, the files are all compressed into a single file, grouping the content of the \verb|output.txt| in a new file, ordering this process by the \verb|rank| of each instance executed, and sent back to the user's download request. In Figure \ref{fig:overview_flow}, we summarize all this process flow.

%% file: sections/architecture.tex
\section{Architecture}
\label{sec:architecture}

This section presents the PESC platform architecture describing its modules and their communication.

\begin{figure}
    \centering
    \includegraphics{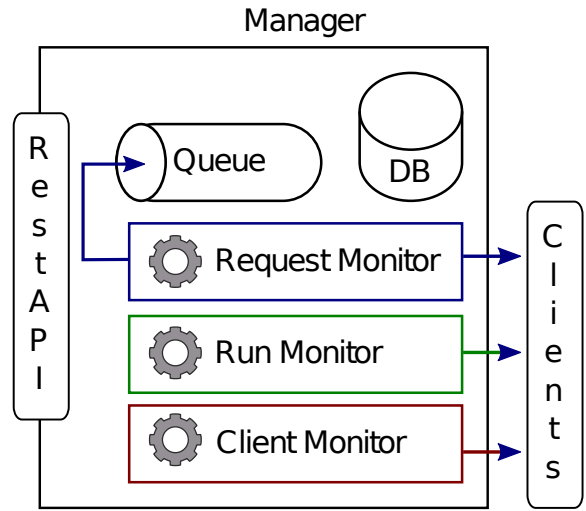}
    \caption{Server Architecture}
    \label{fig:server_architecture}
\end{figure}

\begin{figure}
    \centering
    \includegraphics{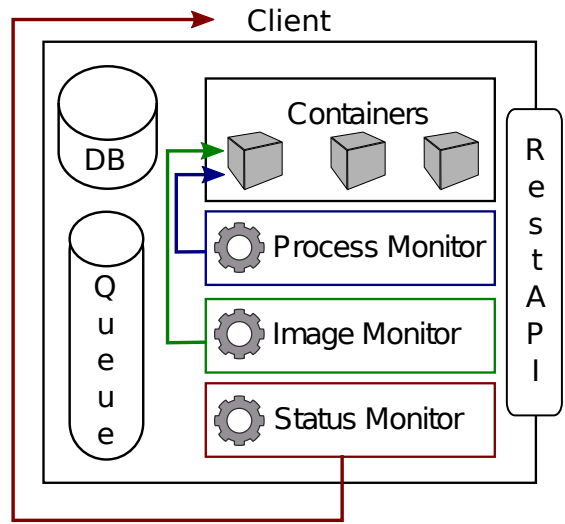}
    \caption{CM Architecture}
    \label{fig:client_architecture}
\end{figure}

As shown in Figure \ref{fig:overview}, the PESC is a system divided into three modules hosted in local infrastructure, the Client, the Manager, and the Frontend modules. The Frontend module is developed in \verb|VueJS Javascript| Framework \citep{street2018complete} and communicates with the Manager through REST API calls, but we will not discuss the Frontend module in this work. Despite being developed using open-source technologies, the PESC platform code is not yet available in the current test phase. After that, however, the platform will be open to the community as an open-source tool. Each module, Client Module and Manager Module, consists of a set of components, as shown in Figure \ref{fig:server_architecture} and Figure \ref{fig:client_architecture}, detailed below. 

\subsection{Manager Module Architecture}
\label{sec:server_architecture}

The Manager Module (MM) is a web server developed with the \verb|Django Python| framework \citep{django_project} that provides an appropriate resource set for the proposed system development. Among these features, we can highlight integration with database (ORM), integrated administration module, user authentication, and access control to system features. Furthermore, integrated with \verb|Django|, the \verb|Django Rest Framework| \citep{django_rest_framework} module has been configured to create communication endpoints with the Client and the Frontend modules through REST API calls. 

The data managed by the platform is stored in a \verb|PostgreSQL| \citep{postgresql12} database. However, \verb|Django| ORM abstraction makes it possible to use other relational databases without changing the system code. In addition, a set of monitors has been developed to manage user \verb|requests| and \verb|clients| communications. Each one of these monitors runs on a different thread and keeps the application state constantly updated.

\subsubsection{Client Monitor}
\label{sec:client_monitor}

The primary function of the Client Monitor is to verify if connected \verb|clients| are available to receive user \verb|processes| to run. Despite being already connected, it is necessary to do this check due to potential network inconsistencies or an unexpected shutdown of one of the \verb|clients|. For example, in this case, if the \verb|client| is reached through the network but does not respond to a REST API call, the Client Monitor will try to restart the Client Module on this machine. This boot possibility is configured on each client via its configuration file.

\subsubsection{Request Monitor}
\label{sec:request_monitor}

Each connected user has a queue used to store their \verb|requests|. The Request Monitor manages these queues and analyzes each \verb|request| which can be set as a single execution or multiple repetitions. In this case, the \verb|request| will continue in the user queue until the total number of \verb|process| instances is forwarded to the available \verb|clients|. \verb|Client| selection is based on the \verb|request| features, for example, whether it needs a GPU or not, and on the workload already distributed to each one.

\subsubsection{Process Run Monitor}
\label{sec:run_monitor}

Each instance of the \verb|process| of a \verb|request| that is sent to a \verb|client| is called a \verb|process run|. The Process Run Monitor checks periodically with the \verb|client|, which received a \verb|process| instance to execute, the status of its execution. \verb|Process runs| that cannot be evaluated must be moved to another \verb|client|, and a cancellation notice must be sent to the current \verb|client|. Offline \verb|clients| will receive the cancellation notification in the upcoming connection with the MM and suspend the execution. Thus, while a \verb|client| is available on the platform, a \verb|process| instance will be directed to it, respecting the limits set in its configuration file, and \verb|requests| should be fully met.

\subsection{Client Module Architecture}
\label{sec:client_architecture}

The Client Module (CM) is an application developed using the \verb|Flask| framework \citep{flask_book} that allows the creation of communication endpoints with the MM and \verb|Docker| containers through REST API calls. The data exchanged with the MM is stored in an \verb|SQLITE3| database \citep{sqlite2020hipp}. However, the SQLALCHEMY package \citep{sqlalchemy} is used, which offers an abstraction layer that allows using other relational databases without the need to make changes to the system code. In addition, a set of monitors has been developed to manage image-building operations and container execution. Each monitor runs on a different thread and continuously updates the application state.

The CM runs as a background service started by an operating system user created for this purpose. Therefore, the CM must balance the use of computer resources with other users who may be using the operating system concurrently. This balance is necessary because \verb|clients| are not exclusively dedicated to the PESC platform.

\subsubsection{Status Monitor}
\label{sec:status_monitor}

Each \verb|client| must inform the MM about the characteristics of its operation. The Status Monitor is responsible for collecting this information to assure the MM that the \verb|client| is in operational condition and it can execute the \verb|processes| received. If the MM does not receive this information, it can move running \verb|processes| to other available \verb|clients|. In addition, the Status Monitor verifies the following information:

\begin{itemize}
    \setlength\itemsep{-0.2em}
    \item Simultaneous use of the machine by other users of the operating system;
    \item The percentage of RAM usage;
    \item The percentage of processor usage;
    \item The percentage of GPU RAM usage, if any.
\end{itemize}

When a user connects to the \verb|client|, the resources allocated from the operating system to the PESC CM are reduced to 10\% not to affect the user experience with the machine. However, when the processor usage percentage reaches 70\%, the \verb|client| informs the MM that it cannot get new processes to run. These percentages are defined in the \verb|client| configuration file.

\subsubsection{Image Monitor}
\label{sec:image_monitor}

Docker containers are built from images that must be pre-built before the \verb|process| instance starts running. The Image Monitor is responsible for monitoring the need to build images and notifying the entire system of this process's progress. A request with the necessary information is sent to the MM if the image is missing on the \verb|client|. Any \verb|process| instance that depends on an image being built will wait for the build to complete successfully before starting execution.

\subsubsection{Process Monitor}
\label{sec:process_monitor}

User-created \verb|processes| run in Docker containers that need an infrastructure of files and directories for correct functioning. The Process Monitor is responsible for monitoring the \verb|process| instance lifecycle and preparing the necessary infrastructure for its operation. The container runs in a detached mode so other containers can be started simultaneously, respecting the operating system resource consumption limit defined in the \verb|client| configuration file. The \verb|process| instance runs in the container through a wrapper that communicates with the \verb|client| through the REST API endpoints.

User \verb|processes| can use the checkpoint directory, informed in the header of received parameters. In case of failure in the container's execution, it will check for any recovery point. If there is a recovery point in the file, the Process Monitor will start a new execution from that point.

%% file: sections/evaluation.tex
\section{Evaluation}
\label{sec:evaluation}

Evaluating the PESC platform is essential to ensure its correct operation and measure the gain in runtime with its adoption.

\subsection{Evaluation Environment}
\label{sec:evalute_environment}

To align with the main objectives of the PESC platform, we created an evaluation environment to emulate a small laboratory with a server and six client computers. Table \ref{tab:software_requirements} presents the software requirements of the evaluation environment, where all computers use Ubuntu 18.04.6 LTS (Bionic Beaver) as the operating system and 500 GB of HD. Table \ref{tab:machine-settings} summarizes the other computer configurations.

\begin{table}[ht]
	\centering
	\caption{Software requirements.}
	\label{tab:software_requirements}
	\begin{tabular}{@{}lll@{}}
		\toprule
		\textbf{Machine Type} & \textbf{Database} & \textbf{Main Packages} \\ \midrule
		Server (Manager)      & PostgreSQL 12     & Django, DRF            \\
		Client                & SQLite 3          & Flask, SQLAlchemy      \\ \bottomrule
	\end{tabular}
\end{table}

\begin{table}[ht]
	\centering
	\caption{Client machine settings.}
	\label{tab:machine-settings}
	\begin{tabular}{@{}lll@{}}
		\toprule
		Machine        & RAM  & System Processor                         \\ \midrule
		Client 1 and 2 & 32GB & Intel(R) Core(TM) i7-2600K CPU @ 3.40GHz \\
		Client 3       & 8GB  & Intel(R) Core(TM) i7-2600 CPU @ 3.40GHz  \\
		Client 4 and 5 & 32GB & Intel(R) Core(TM) i7-6700 CPU @ 3.40GHz  \\
		Client 6       & 16GB & Intel(R) Core(TM) i7-8700 CPU @ 3.20GHz  \\ \bottomrule
	\end{tabular}
\end{table}

\subsection{Evaluation Execution}
\label{sec:evalute_experiment}

To evaluate the PESC platform, we created several scenarios, and in each one, the platform must execute the \verb|requests| created by the user and finalize the expected results. We present each scenario in more detail below:

\begin{itemize}    
    \setlength\itemsep{-0.2em}
    \item Run a simple code;
    \item Repeat the simple code N times;
    \item Run a specialized code;
    \item Run a specialized code with parallelization;
    \item Recover after failure on Client Module (CM) or Manager Module (MM);
    \item Run a code that uses a framework with native parallel features.
\end{itemize}

\subsubsection{Scenario 1 - Run a simple code}
\label{sec:evalute_experiment1}

In this scenario, a user must run a simple code made in \verb|Python| using only built-in functions. The code is a Gaussian random number generator that generates 10.000.000 numbers. The user will use an available \verb|domain| (Simple \verb|Python|) from the \verb|store| that will abstract all the container infrastructure knowledge needed and not make any changes to their code. This scenario is the most basic use for the PESC platform, and Algorithm \ref{alg:scenario_1} shows a portion of the user pseudocode.

\begin{algorithm}[H]
\begin{algorithmic}
\State $x \gets 0$
\For{$rank=0$ to $10000000$}
\State $u1, x \gets get\_uniform(x)$
\State $u2, x \gets get\_uniform(x)$
\State $z1, z2 \gets get\_gaussian(u1, u2)$
\State \textbf{print} \texttt{rank+``: ''+z1 +``,''+z2}
\EndFor
\end{algorithmic}
\caption{Scenario 1 - Validation code}
\label{alg:scenario_1}
\end{algorithm}

This scenario requires the assignment of the user code to a new \verb|process| (Validation1\_1). To run this new \verb|process|, the user creates a new \verb|request| with the values of the following parameters:

\begin{itemize}    
    \item Domain: Simple \verb|Python|
    \item Process: Validation1\_1
    \item Repetitions: 1
    \item Rooms: Public
\end{itemize}

The PESC platform selects a \verb|client| to distribute and run this \verb|process|, which runs as expected on this \verb|client| and returns its output in a zipped file to the user. We also validate the execution of an \verb|R| script and a \verb|Java| code version for this scenario. In these cases, the user must create the \verb|domain| for each technology or select it at the \verb|store|. In addition, to execute code written in \verb|Java|, the user must upload the \verb|process| file in executable JAR format. Both simulations were run and returned the expected output files.

\subsubsection{Scenario 2 - Repeat the simple code N times}
\label{sec:evalute_experiment2}

Suppose the user needs to generate more random numbers. In that case, it could be done without any changes to their code, setting a value greater than one on the Repetitions parameter, as listed below. 

\begin{itemize}    
    \setlength\itemsep{-0.2em}
    \item Domain: Simple \verb|Python|
    \item Process: Validation1\_1
    \item Repetitions: 10
    \item Rooms: Public
\end{itemize}

The PESC will distribute each instance of this \verb|process| to the available \verb|clients| and return all generated output in a single zipped file attached to the \verb|request|. In addition, an individual zipped file is also attached to each \verb|process run|.

\subsubsection{Scenario 3 - Run a specialized code}
\label{sec:evalute_experiment3}

In this scenario, a user needs to run a simulation written in \verb|Python| using Scikit-learn \citep{pedregosa2011scikit} library features but cannot make changes to the machine operating system to install the necessary software. In this case, using the PESC platform will be an abstraction for containers. To sort the MNIST digit dataset, converted to CSV format, the user coded a $k$-nearest neighbors (kNN) algorithm \citep{10.1145/3459665}, and Algorithm \ref{alg:scenario_3} shows a portion of the user pseudocode.

\begin{algorithm}[H]
\begin{algorithmic}
\State $train \gets read\_csv('./mnist\_test.csv')$
\State $test \gets read\_csv('./mnist\_train.csv')$
\For{$rank=0$ to $10$}
\State $model \gets knn(n=rank)$
\State $model.fit(train.x, train.y)$
\State $accuracy \gets model.score(test.x, test.y)$
\State \textbf{print} \texttt{``k=''+rank+``==>''+accuracy}
\EndFor
\end{algorithmic}
\caption{Scenario 2 - Validation code}
\label{alg:scenario_3}
\end{algorithm}

The requirements needed for this scenario are the Dockerfile, requirements.txt, user code, and the CSV file with a digit base. The user creates the \verb|domain| (Scenario\_3) with the first two files or uses the \verb|Python| environment in the \verb|store|, sets the necessary packages to customize it, and creates the \verb|process| (Validation\_3) with the user code. We made the test and training databases in a process before this execution, and they will be uploaded to the PESC platform as \verb|shared files| and will be available to the code on the same directory path. This way, other validation scenarios presented in this Section can use the \verb|shared files| without the need to transfer the files again for \verb|clients| that have already participated in the execution of this scenario. Finally, the user creates the \verb|request| with the following parameters:

\begin{itemize}    
    \item Domain: Scenario\_3
    \item Process: Validation\_3
    \item Repetitions: 1
    \item Shared Files: minst\_test, mnist\_train
    \item Rooms: Public
\end{itemize}

The code runs as expected and returns the output file with the result shown in Listing \ref{lst:lst2} with the accuracy found for each value of k:

\begin{lstlisting}[caption=Scenario 3 - Output, label={lst:lst2}, captionpos=b]
k=1==>0.9416833333333333
k=2==>0.93195
k=3==>0.9428333333333333
k=4==>0.94115
k=5==>0.9425166666666667
k=6==>0.9404333333333333
k=7==>0.9401333333333334
k=8==>0.93905
k=9==>0.9380166666666667
k=10==>0.9371666666666667
\end{lstlisting}

\subsubsection{Scenario 4 - Parallelize the specialized code} 
\label{sec:evalute_experiment4}

\begin{figure}[H]
    \centering
    \includegraphics{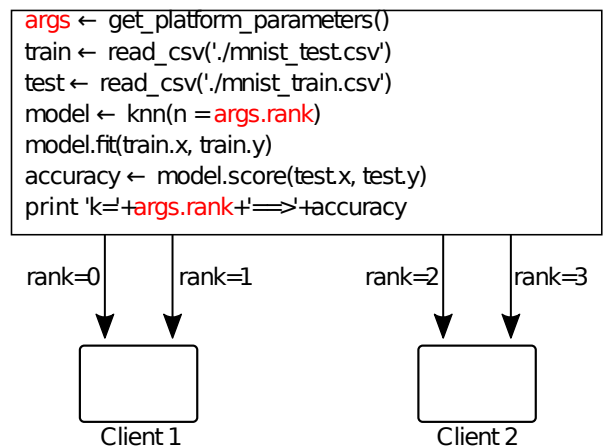}
    \caption{Ranks distribution}
    \label{fig:ranks_distribution}
\end{figure}

The PESC platform executes user code by passing a set of parameters through a command line call, as presented in Section \ref{sec:system_overview}. Among these parameters is the ID of the running instance, called rank, which is a value starting at 0. The user can use this parameter to transform a sequential loop into a set of parallel executions. In this case, instead of a single instance with a loop of $N$ repetitions, we can have $N$ instances of the program where each one executes the loop's contents only once, as shown in Figure \ref{fig:ranks_distribution}. In this way, each instance would represent an iteration of the loop through the rank variable.

In this scenario, the user changes the code used in the previous one to execute each value of $k$ in a different \verb|process| instance. This way, a sequential code will be executed in parallel with minimal changes to the user code, as shown in Algorithm \ref{alg:scenario_4}. In this case, the user uses the platform more optimally than in the previous scenario.

\begin{algorithm}[H]
\begin{algorithmic}
\State $args \gets get\_platform\_parameters()$
\State $train \gets read\_csv('./mnist\_test.csv')$
\State $test \gets read\_csv('./mnist\_train.csv')$
\State $model \gets knn(n=args.rank)$
\Comment{The loop started here}
\State $model.fit(train.x, train.y)$
\State $accuracy \gets model.score(test.x, test.y)$
\State \textbf{print} \texttt{``k=''+rank+``==>''+accuracy}
\end{algorithmic}
\caption{Scenario 4 - Validation code}
\label{alg:scenario_4}
\end{algorithm}

The user creates a new \verb|request| with the values of the following parameters:

\begin{itemize}    
    \setlength\itemsep{-0.2em}
    \item Domain: Scenario\_3
    \item Process: Validation\_4
    \item Repetitions: 10
    \item Shared Files: minst\_test, mnist\_train
    \item Rooms: Public
\end{itemize}

The requirements needed for this scenario are the same as the last one. The code runs as expected and returns the output files from each \verb|client| with the result found for the $k$ value. The output files are concatenated into a single file to facilitate the analysis of the results by the user.

\begin{figure}[H]
    \centering
    \includegraphics[width=0.8\linewidth]{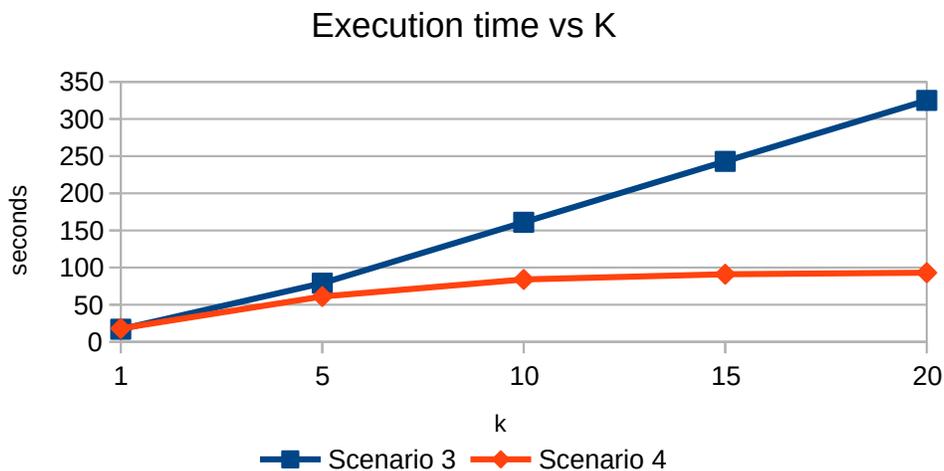}
    \caption{Comparison between sequential and parallel scenarios}
    \label{fig:scenario_comparsion}
\end{figure}

Validation\_3 and Validation\_4 processes were executed several times with $k$ = 1, 5, 10, 15, and 20 to validate the gain using the PESC platform. The execution times were compared, as shown in Tables \ref{tab:scenario3_execution_time} and \ref{tab:scenario4_execution_time}. We did not consider the time used by the \verb|clients| to build the Docker image. Figure \ref{fig:scenario_comparsion} shows the gain from the changes made to the code for this scenario.

\begin{table}[ht]
\centering
\caption{Scenario 3 execution time}
\label{tab:scenario3_execution_time}
\begin{tabular}{@{}lccr@{}}
    \toprule
    \textbf{K} & \textbf{Start Time} & \textbf{End Time} & \textbf{Seconds} \\
    \midrule
    1          & 13:48:42            & 13:48:59          & 17               \\
    5          & 13:50:25            & 13:51:44          & 79               \\
    10         & 13:56:19            & 13:59:00          & 161              \\
    15         & 14:08:03            & 14:12:06          & 243              \\
    20         & 14:13:57            & 14:19:22          & 325              \\
    \bottomrule
\end{tabular}
\end{table}

\begin{table}[H]
\centering
\caption{Scenario 4 execution time}
\label{tab:scenario4_execution_time}
\begin{tabular}{@{}lccr@{}}
    \toprule
    \textbf{K} & \textbf{Start Time} & \textbf{End Time} & \textbf{Seconds} \\
    \midrule
    1          & 14:32:02            & 14:32:20          & 18               \\
    5          & 14:33:55            & 14:34:56          & 61               \\
    10         & 14:39:23            & 14:40:47          & 84               \\
    15         & 14:48:04            & 14:49:35          & 91               \\
    20         & 14:53:26            & 14:54:59          & 93               \\ 
    \bottomrule
\end{tabular}
\end{table}

\subsubsection{Scenario 5 - Recover after failure on Client Module (CM) or Manager Module (MM)}
\label{sec:evalute_experiment5}

In this scenario, we will repeat the same \verb|request| created in the previous one but disconnect two \verb|clients| from the MM after receiving the \verb|process| instance to execute. If any communication failure occurs with one of the \verb|clients|, the MM cancels the sending process and redistributes it to another \verb|client|. The code works as expected, and we can verify the redistributed processes in the database table presented in Listing \ref{lst:lst4}. The user can verify this behavior through the web interface.

\begin{lstlisting}[caption=Scenario 5 - Database query, label={lst:lst4}, captionpos=b]
  id   | rank | client_id | status |   obs        
-------+------+-----------+--------+-----------
 23547 |    0 |        11 |      3 | Sucess
 23548 |    1 |        12 |      3 | Sucess
 23549 |    2 |         8 |      3 | Sucess
 23550 |    3 |        10 |      5 | Canceled
 23557 |    3 |        12 |      3 | Sucess
 23558 |    4 |         8 |      3 | Sucess
 23551 |    4 |         7 |      5 | Canceled
 23552 |    5 |        12 |      3 | Sucess
 23553 |    6 |         8 |      3 | Sucess
 23554 |    7 |         7 |      5 | Canceled
 23559 |    7 |        11 |      3 | Sucess
 23555 |    8 |        11 |      3 | Sucess
 23556 |    9 |        11 |      3 | Sucess
\end{lstlisting}

As we can see, the \verb|process| executed with $id=23550$ and $rank=3$ was canceled on the \verb|client| with $id=10$, but the \verb|client| with $id=12$ completed the same rank value. The same happened with \verb|processes| with ids equal to 23558 and 23554. This redistribution guarantees that as long as \verb|clients| are connected and resources are available, the PESC platform will successfully execute the user code and complete the \verb|request| without any user iteration. This behavior is also one of the main objectives of the PESC platform.

In case of MM failure, \verb|clients| continue executing the received \verb|process| instances. They will send the execution status when the MM is back and make a new connection between them. This process does not change the initial flow of execution after the distribution of the \verb|process| instances.

\subsubsection{Scenario 6 - Run a code that uses a framework with native parallel features}
\label{sec:evalute_experiment6}

Some frameworks offer native features for parallel programs, such as the Pytorch Distributed RPC Framework (DRF) \citep{pytorch_rpc}. In this scenario, we execute a \verb|process| where each instance will assume different roles due to the requirements of this resource. This scenario was built based on code from the official PyTorch tutorial \citep{pytorchwebsite}, where distributed training was made using the \texttt{torch.distributed.rpc} package. Here, only distributed reinforcement learning using RPC and RRef code was used. In this scenario, a \verb|client| will play the agent role, coordinating the sending of data to other \verb|clients|, that will play the observer role. The agent has a central role, and the observers need to identify this \verb|client| to establish communication between them. The PESC platform informs the IP address and the port created for the \verb|client| that received $rank=0$ for each \verb|process| instance. In this way, the other instances, $rank>0$, can communicate with this instance and meet the requirements of the resource used in the Pytorch DRF. The user code checks the rank parameter, and the necessary adjustments are made, as shown in the Algorithm \ref{alg:scenario_6}. Finally, the user creates the \verb|request| with the following parameters:

\begin{itemize}    
    \setlength\itemsep{-0.2em}
    \item Domain: Scenario\_6
    \item Process: Validation\_6
    \item Parallel: True
    \item Repetitions: 3
    \item Rooms: Public
\end{itemize}

\begin{algorithm}[H]
\begin{algorithmic}
\State $args \gets get\_platform\_parameters()$
\State $...$
\If{$args.rank == 0$} 
\Comment{rank=0 is the agent}
\State $name \gets agent\_name$
\State $rpc.init\_rpc(name, args.rank, args.world\_size)$
\State $agent \gets Agent(args.world\_size, gamma)$
\Repeat
\State $agent.run\_episode()$
\State $last\_reward \gets agent.finish\_episode()$
\Until{$agt.running\_reward>agt.reward\_threshold$}
\State print ``Solved!''
\Else
\Comment{other ranks are the observer}
\State $name \gets observer\_name+args.rank$
\State $rpc.init\_rpc(name, args.rank, args.world\_size)$
\State observers passively waiting for agent instructions 
\EndIf
\State block until all rpcs finish, and shut down the RPC instance
\State $rpc.shutdown()$
\end{algorithmic}
\caption{Scenario 6 - Validation code}
\label{alg:scenario_6}
\end{algorithm}

The code runs as expected, where the repetitions parameter in the \verb|request| form determines the number of instances that \verb|clients| will execute. The Parallel parameter prevents \verb|processes| from starting unsynchronized. This information will make all \verb|clients| that received an instance of the \verb|process| wait for a signal from the MM before beginning execution. This option helps minimize resources being allocated to containers that are not running their \verb|processes|.

%% file: sections/experimental_results.tex
\section{Experimental Results - A Real Case}
\label{sec:experimental_results}

Although the validation tests presented the expected results, a use case was used to evaluate the platform's value for a real user. This use case is about the simulation of a quantum search procedure that uses the lackadaisical quantum walk algorithm in an $n$-dimensional hypercube to search for multiple solutions\citep{souza2021lackadaisical}.

The computational cost of simulating a quantum system tends to grow exponentially and requires a large amount of memory and computer runtime \citep{acheson2001feynman}. As a result, a classical system may not efficiently simulate quantum systems \citep{feynman1982simulating}.


This quantum walk algorithm is characterized by using self-loops in each vertex of the structure used as a representation of the search space, which in this case was the hypercube. The simulation was divided into three scenarios that depended on the type of vertex marked as a solution: Non-adjacent vertices in the first scenario, adjacent vertices in the second scenario, and both adjacent and non-adjacent vertices in the third scenario.

Another feature of the lackadaisical quantum walk is the dependence on the weight value for the self-loop \citep{de2020impacts,souza2021lackadaisical,decarvalho2021applying}. This weight value adjusts the probability of a walker staying at the vertex. Experiments were performed in each of the three scenarios using four weight values for the self-loop. It was necessary to obtain the average behavior of the quantum walk based on the relative position of the non-adjacent marked vertices. Therefore, in scenarios with non-adjacent marked vertices, one hundred simulations were performed for each set of marked vertices.

This way, twelve hundred simulations were performed for the scenario where the marked vertices are only non-adjacent. Eleven hundred simulations were performed for the scenario where the marked vertices are adjacent and non-adjacent. Only twelve simulations were performed for the scenario where the marked vertices are adjacent. Each of the simulations was performed in one rank. Each rank was responsible for executing thirty lackadaisical quantum walks ranging from one to two hundred iterations. At the end of the process, the maximum probability of success of thirty walks was returned in each rank. To run the scenario where the marked vertices are adjacent and non-adjacent, and the self-loop weight is three, the user creates a new \verb|request| with the values of the following parameters:

\begin{itemize}
    \setlength\itemsep{-0.5em}
    \item Domain: QuantumWalk
    \item Process: adjacent-non-adjacent
    \item Parallel: False
    \item Repetitions: 1200
    \item Parameters: 3
    \item Rooms: Public
\end{itemize}

In this case, the \verb|shared file| was not used, but the process was created with the vertices files and the user code as a zip file. Using the platform simplified the simulation execution process, as it manages the status and life cycle of the process. Four registered \verb|clients| participated in this execution. Thus, running the simulations in the same laboratory environment used for the validation tests was possible.

\verb|Client| nodes are not exclusive to the PESC platform. They may compete for computational resources with other systems, and the \verb|client's| settings are different, as shown in Table \ref{tab:pesc-distribution}. Therefore, the distribution of instances is not balanced, as shown in the average duration column in Table \ref{tab:pesc-distribution}.

\begin{table}[H]
\centering
\caption{Instances distribution by clients.}
\label{tab:pesc-distribution}
\begin{tabular}{@{}llr@{}}
    \toprule
    \textbf{Machine} & \multicolumn{1}{l}{\textbf{Avg Duration}} & \multicolumn{1}{l}{\textbf{Count}} \\
    \midrule
    Client 1         & 01:20:38.42278 & 207                  \\
    Client 2         & 01:22:21.366742 & 202                 \\
    Client 3         & 01:01:13.74276 & 224                  \\
    Client 6         & 00:31:24.892872 & 567                 \\
    \bottomrule
\end{tabular}
\end{table}

Considering that the lowest average time for the execution of the instances was 00:31:24s, if all 1200 sequential repetitions, original code format, were executed on this \verb|client|, the execution time would be approximately 600 hours. Therefore, the total execution time on the PESC platform was approximately 12:39:14s, considering the moment when the first instance starts its execution and the last one ends.

%% file: sections/conclusions.tex
\section{Conclusions}
\label{sec:conclusions}

Using computational infrastructures as a computational grid for load distribution of parallel execution has already proven to be a viable option for developing scientific research. We present a general-purpose platform focused on the actors involved in the process, users, and IT teams, that is, simplicity of use and ease of maintenance and configuration of the platform. The abstraction of the technologies necessary to create this platform allows programs developed by any user and in any programming language to be executed without practically any modification. It also provides code initially designed for a sequential execution to run in parallel with minimal user adjustments. Furthermore, the use of idle resources of the institution allows teaching and research institutions to make more optimal use of these resources without the cost of acquiring new equipment for this purpose. We show through several scenarios that the platform remains simple even for a more advanced use case that depends on specific features of languages or frameworks. We also show that a simulation that intensively used computational resources in its sequence version was easily adapted to a parallel version and presented significant performance gains even when using computational resources in everyday use.